# Access control management for e-Healthcare in cloud environment

Lili Sun [1,*], Jianming Yong [2] and Jeffrey Soar [2]

[1] Centre for Systems Biology, University of Southern Queensland, Toowoomba, QLD4350, Australia
[2] School of Management and Enterprise, University of Southern Queensland, Toowoomba, QLD4350, Australia

## Abstract

Data outsourcing is a major component for cloud computing that allows data owners to distribute resources to external services for users and organizations who can apply the resources. A crucial problem for owners is how to make sure their sensitive information accessed by legitimate users only using the trusted services but not authorized to read the actual information. With the increased development of cloud computing, it brings challenges for data security and access control when outsourcing users' data and sharing sensitive data in cloud environment since it is not within the same trusted domain as data owners'. Access control policies have become an important issue in the security filed in cloud computing. Semantic web technologies represent much richer forms of relationships among users, resources and actions among different web applications such as clouding computing. However, Semantic web applications pose new requirements for security mechanisms especially in the access control models. This paper addresses existing access control methods and presents a semantic based access control model which considers semantic relations among different entities in cloud computing environment. We have enriched the research for semantic web technology with role-based access control that is able to be applied in the field of medical information system or e-Healthcare system. This work shows how the semantic web technology provides efficient solutions for the management of complex and distributed data in heterogeneous systems, and it can be used in the medical information systems as well.

## 1. Introduction

Cloud computing is the delivery of computing and storage capacity as a service to a community of end-recipients. With the advanced technology of network and software system, cloud computing is becoming a new variation of traditional distributed computing and grid computing [9, 21, 30]. Cloud computing allows users to put all data and services into cloud and gets all kinds of services from cloud such as software as a service and database as a service. That makes remote collaboration between organisers easier, especially in the e-healthcare systems. With the increasing development of cloud computing technologies, applications and storage of information will be significant changed in cloud computing environment. As a result, more and more businesses will be moved into the cloud in the future.

However, the development of cloud computing is still facing enormous security challenges due to the existing secure technology is not able to directly apply to cloud environment. Cloud computing, with a great flexibility and ease of use, makes the safety of data and applications becoming one of the biggest problems [9]. In fact, applications and information using of cloud hosting have the risk of loss of data or illegal access. Therefore, it needs to have appropriate permissions when users to access the applications or data. At present, there are already many security specification and technologies [5, 15]. However, access control is a critical component in many environments, such as cloud computing. Cloud computing cannot apply the traditional access control models to achieve access control because of its characters.

*Corresponding author. Email: sun@usq.edu.au

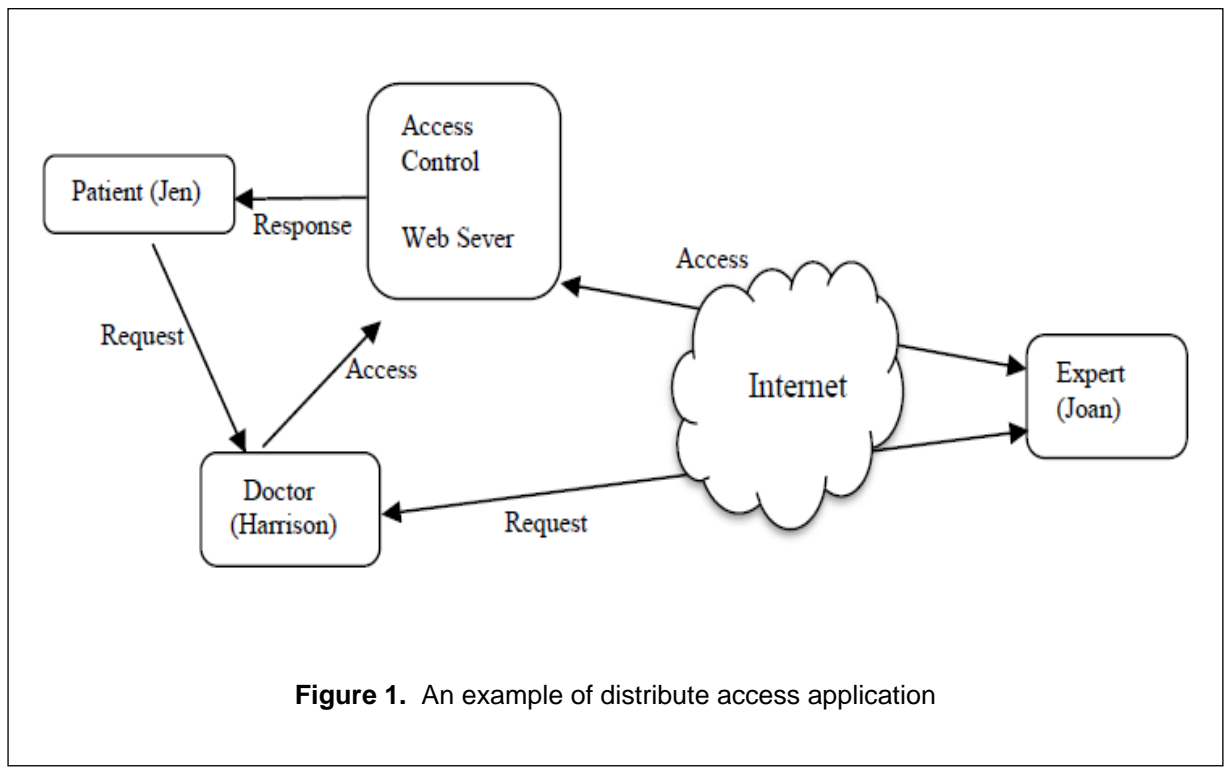

**Figure 1.** An example of distribute access application

Traditional access control methods cannot provide the semantic interoperability for many software network environments since in the distributed environment, access control policies are used among various computers and servers in various places.

In cloud computing environment, one organization seeking to share information with authorized personnel from other organizations has to deal with semantic heterogeneity across the information sources in these organizations [14, 24, 28, 31]. An example is a medical centre where a patient is under the treating of a medical centre doctor. Assume the doctor needs to consult an expert who works in a hospital, since some abnormality detected in the captured patient's physiological parameters such as temperature, heartbeat, blood oxygen, blood pressure, etc. The consulting expert receives physiological data through his laptop or PC. A significant challenge in this situation is to control who can see the data and what conditions. Due to the ability of remotely accessing the patients vital signs it is important that this information not be remotely accessible at all times the patient in the hospital, but perhaps only under particular situations like the expert worked more than five years.

There is a scenario is shown in Figure 1. In this scenario, a patient, Jen, is at a medical centre with a doctor, Harrison. There is a web service attached to Jen in order for Harrison to monitor physiological data. Harrison requires further assistant from an expert Joan. Joan can access the web service via his browser to receive and analyse patient's physiological data and give some advices to doctor Harrison. The target is to permit the sharing of resources among the participants, whether or not they are at the same site. For a conventional access control model, this can be easily done within the same site, such as only a doctor in medical centre. Also the differences in the vocabulary used by the hospitals have to be resolved before patients information can be shared meaningfully among the different hospitals. To overcome these challenges, there is a need for semantic aware access control systems consistent with the semantic data models under the semantic web in the cloud computing environment.

E-healthcare informatics is growing need for healthcare providers to have effective healthcare services to consumers and provide health information to guide consumers accessing the information which they need. It is an application for web-based systems to share and deliver information across the Internet. Private health information once confirmed to these local networks, they could be migrated onto the Internet. Patients are able to check their own health records and buy prescription medicine online. In order to offer "in the cloud" products and services to patients, doctors, and institutions, the insurer companies also need to accelerate access to these records. The healthcare information system provides many advantages when used for improved access, collaboration and data sharing among healthcare providers, patients, and researchers. Therefore, considering the highly personal and potentially destructive nature of medical data, it comes with significant risks to the confidentiality, integrity, and availability of such information. This coming explosion of information will be stored in massive data centres around the world and will provide access to healthcare records for patients, insurers, doctors, pharmacies, and institutions. According to the Certification Commission for Healthcare

Information Technology (CCHIT), more than 300 vendors currently offer some variance of electronic medical records — some "in the cloud," some locally, and some in both [2].

Unlike other applications, medical information system requires a much more effective access approach due to the fact that collecting countrywide, worldwide information system is paramount for health organizations, governments, and world-wide agencies [1, 4]. However, it is getting hard to extract useful information from the enormous amount of data that is being collected in the medical information system or e-Health system. Traditional healthcare datasets housed within hospitals are mainly governed by a set of privacy regulations that determine the aim and scope of the registration, the type of data, the rights of data subjects as well as access rights [1, 8]. E-healthcare services require that the systemic use of protected health information pose additional potential threats to the security, privacy, and confidentiality of e-patient information. For example, one published survey reported that in the United States, some people do not file insurance claims or see health service providers for fear that disclosure of their health information may hurt their job prospects or ability to obtain insurance coverage [1, 17]. Here, semantic web technologies could play a greater important role as we will present in this paper. Semantic web technologies are based knowledge bases. It aims to first annotate data via common ontology, then interlink data on the web and allow one to query this web of knowledge. It is possible to apply this approach to the medical information system integration. This paper proposes a semantic access control approach in e-Health system, which applies semantic web technology to access control method. This research provided a new application for access control in cloud computing environments.

Having data on web defined and linked in a way that it can be used for more effective discovery, integration and reuse across various applications. However, the shift data from centralized to distributed environments such as semantic web poses new security requirements especially in the field of access control. Access control is a mechanism that allows owners of resources define, manage and enforce access conditions applicable to each resource [9]. In the centralized systems, the same entity is responsible for the assignment of attributes to clients and access control [10]. All the information required to analyse and manage are locally in the same system where the resources reside. The relevant problems are the lack of interoperability for the open distributed systems. Traditional access control model like role based access control (RBAC) is considered a mature and flexible technology, the access control policies depend on the roles that users play within the system but they do not consider the rich semantic relations in the data model under semantic web [5, 11, 29]. Semantic access control (SAC) model provides an appropriate solution, especially for heterogeneous, distributes and large environments such as cloud computing. It complements the use of attributes with the use of metadata to represent the semantics different elements.

Ontology defines a common vocabulary for people who need to share information in a domain [12]. Ontology is helpful to construct authorization policy within the scope of whole cloud computing environment sharing common understanding of the structure of information among people. For example, suppose several different web sites contain medical information or provide e-healthcare services. If these web sites share and publish the same underlying ontology of the terms they all use, then researchers can extract and aggregate information from these different sites. The researchers can use this aggregated information to answer user queries or as input data to other applications. Ontologies are used for modeling the entities along with their semantic interrelations in four domains of access control, namely subject domain, object domain, action domain and attributes domain [14]. In this paper, we use the semantic scopes of subject, objects, actions, and attributes to define the relations used in ontologies. We present a semantic based access control model that authenticates users based on ontologies with e-Healthcare system.

The remainder of this paper is organized as follows: Section 2 provides a brief overview semantic web and access control ontology system in e-Healthcare services. In Section 3 and 4 present semantic access control policy and semantic access control model in cloud computing. Section 5 illustrates the implementation with XACML. Section 6 compares the work in the paper to others related works. Finally, Section 7 concludes the paper.

## 2. Semantic web and access control ontology system

### 2.1. RBAC and Semantic Web Technologies

RBAC model [18, 28] includes set of three basic elements: users, roles and permissions as shown in Figure 2. RBAC involves individual users being associated with roles as well as roles being associated with permissions (each permission is a pair of objects and operations). As such, a role is used to associate users and permissions. A user in this model is a human being. A role is a job function or job title within the organization associated with authority and responsibility. Permission is an approval of a particular operation to be performed on one or more objects. Access control policies specify user's permissions to specific system resources through relationships between user's roles and permissions. The relationships between users and roles and between roles and permissions are many-to-many (i.e. permission can be associated with one or more roles, and a role can be associated with one or more permissions).

Semantic web is the extension of current web which gives information a well-defined meaning, better enabling computers and people to work in co-operation [24]. The semantic web provides a framework for dynamic, distributed and extensible structured knowledge founded on formal logic [16]. Semantic web techniques, particularly ontologies, facilitate web services with machine understandable semantics. Ontologies providing new features namely automatic composition, simulation and discovery of web services [10]. Ontologies are essential for semantic interoperability and advanced information processing as web services allow computation over the web. Using ontologies to describe relationships between data is increasingly used in information and knowledge management. Ontology is capable of describing concepts that exist in certain domain and relationships among them. The domain concepts, such as personal information, medical information, are captured and formalized with ontology. Figure 3 shows a part of E-health service ontology. The ovals show subjects and objects (concepts and individuals) and labels on the directed arcs shows actions.

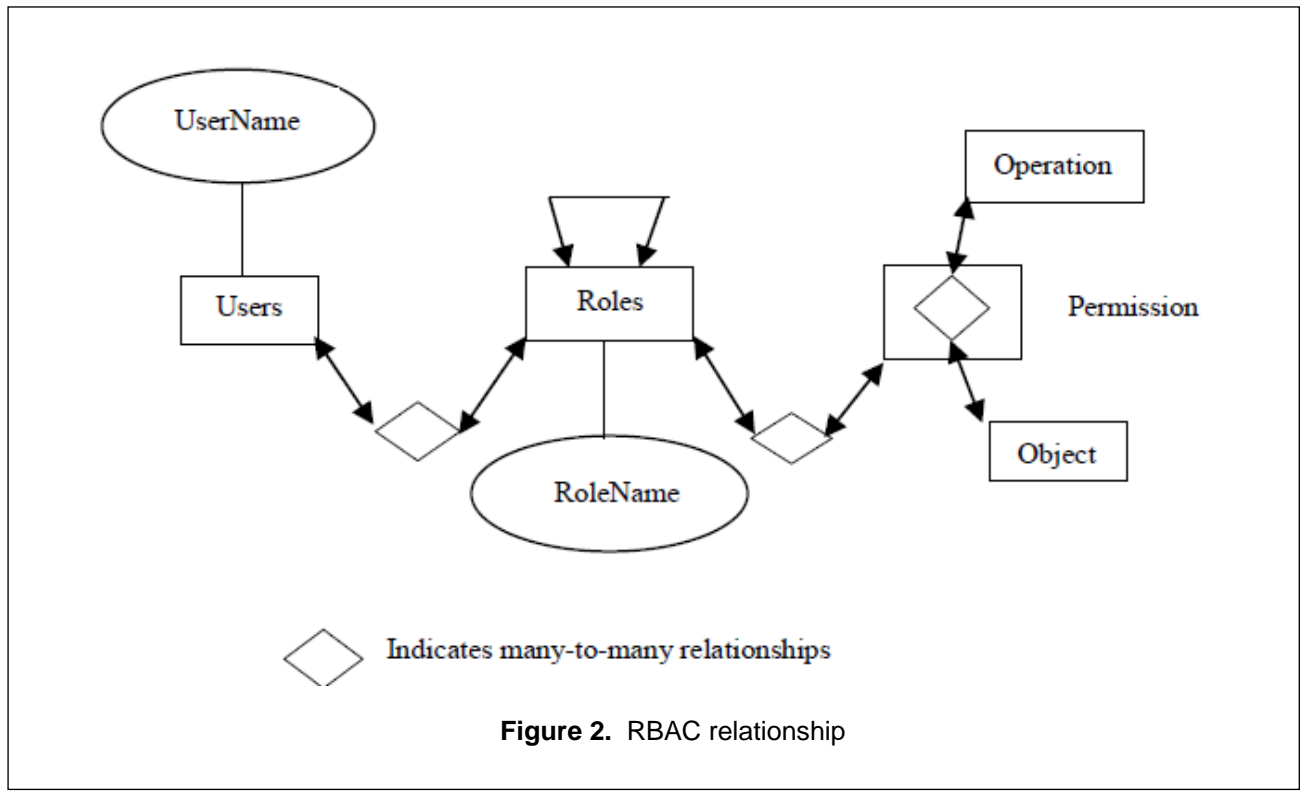

**Figure 2.** RBAC relationship

## 2.2. Semantic access control

Semantic access control (SAC) in this paper is based upon RBAC. In this paper, we use ontologies for the RBAC security model and implement access control system in semantic web environment. Our goal is to request and extend the RBAC model using semantic web technologies. RBAC involves additional effort from the host organizations in deciding which roles or users from remote organizations should have access to which object.

Based on RBAC, a semantic authorization rule has the following definition [14]:

There is a triple $(role, object, act)_{onto}$, where, *role* is the role of the user who issues the request; *object* is the object defined in *onto* to which the user requests to access; *act* is the action to be executed on the object; *onto* is the ontology to which the authorization associates.

Figure 4 shows the graph of the semantic access control model based on RBAC. Role acts an intermediary for assigning permissions (objects and operations) to users which greatly simplifies authorization administration. Action is a partial, or class that represents an action that can be performed by a user on a resource. Resource is a defined class, representing the authorization objects. We can identify all the objects that have been treated like a resource in the domain ontology. For example, triple $(r, o, read)_{onto}$ indicates that users assign to role $r$ can read information form object $o$ defined in ontology *onto*.

In this article, access control ontology system is designed to provide the common understandable semantic basis for access control in cloud computing environments. In SAC, the subjects and objects possess a set of attributes and access control to resources is based on the specification of a set of attributes. Like most of the other access control systems, access control ontology system makes its decisions on four domains: Subject, Object, Action and Attributes. By modelling the access control domains using ontologies, SAC provides a set of ontologies: Subjects–Ontology (SO), Objects–Ontology (OO), Actions–Ontology (AO), and Attributes-Ontology (AtO) [25]. SO is subject ontology where subjects require access to object. OO is object ontology where objects are accessed and /or modified. Figure 3 shows subject-ontology and object ontology which is based on e-Healthcare system. Actions depend on the type of the actions that subjects aim to execute on an object. Each action type is a concept in the ontology and actions are individuals of the concept defined in AO [25]. Figure 5 shows an example of action ontology. AtO is the attribute ontology which can be used to the attribute of the subjects, objects and actions. For example, an authorization rule in an access control ontology with the form of (s, o, a) in which $s$ is an entity in SO, $o$ is an entity defined in OO, and $a$ is an action defined in AO. In the other words, an access rule determines whether a subject which presents a subject $s$ can have the access right $a$ on object $o$ or not.

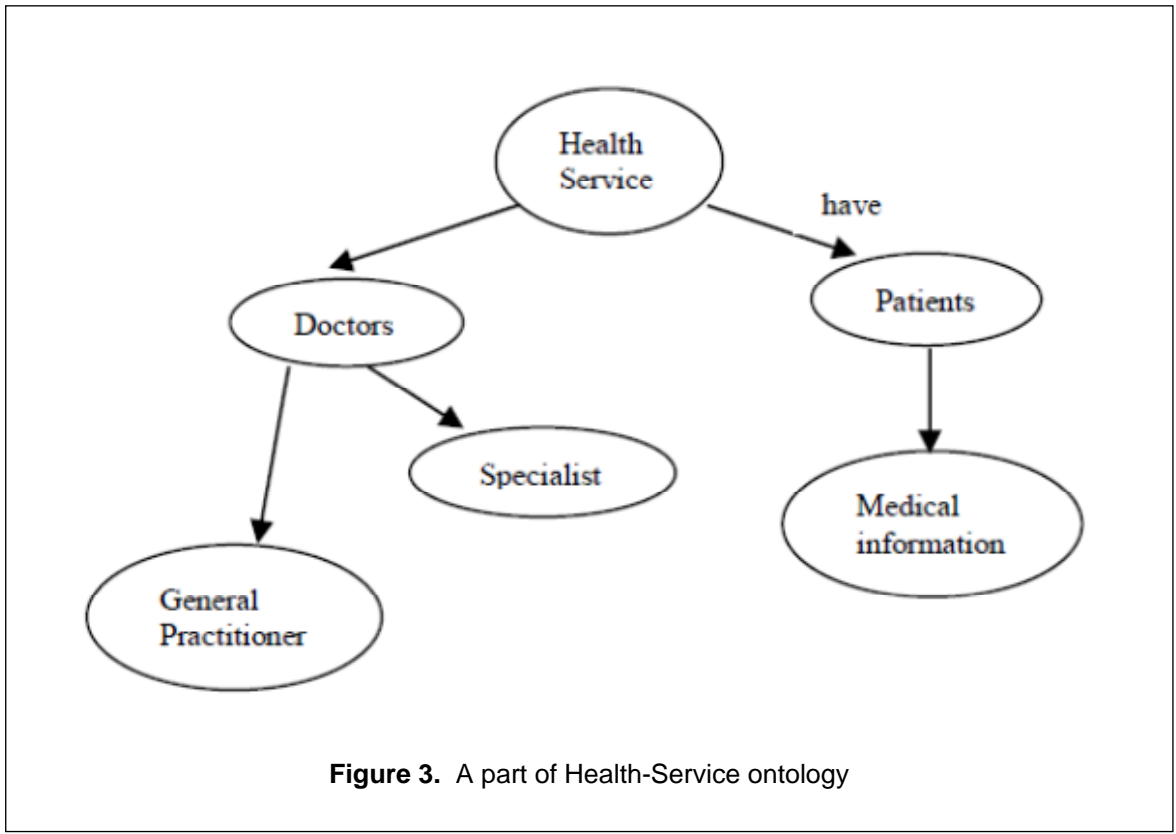

**Figure 3.** A part of Health-Service ontology

The role in subject ontology may be used as a subject attribute, depending by the attributes provided a user is assigned to a certain role policy set [7]. Other attributes can also be associated with the subject in order to achieve fine-grained access control. If a subject is assigned to a role, it cannot access the resources directly. Meanwhile, the roles are organized in a hierarchy. If a role $r1$ inherits from role $r2$ in the hierarchy, a user with $r2$ has all the access rights of $r1$.

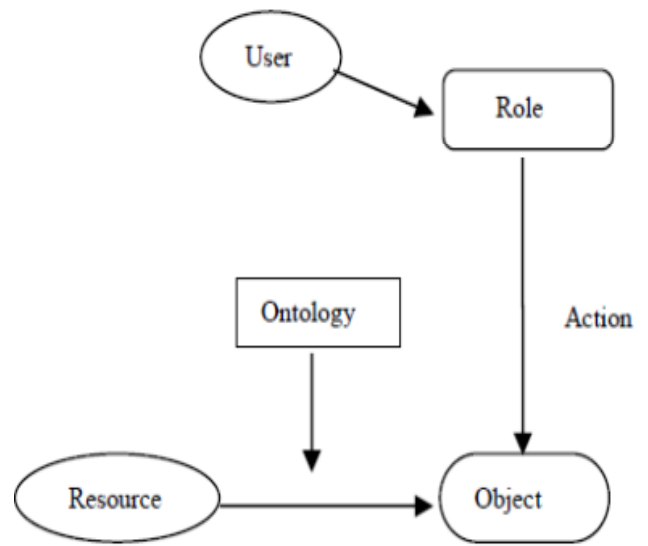

**Figure 4.** Semantic access control model

# 3. Semantic access control policy in cloud computing

In the distributed computing environment, the access control method has changed from the centralized management into a distributed management approach. There has been policy Markup language, such as XACML, to support description and management of distributed policies. In cloud computing environment, as the development of distributed computing, the same access control policy may be deployed and implanted in many points of the whole or a part of the security domain [14]. SAC has been implemented on the basis of a language to specify the access control criteria and the semantic integration of external authorization entities [12]. In an ontology-based semantic access control policy language, we may use subjects, objects, actions and attributes variables as the basic semantic element and some syntax elements such as purposes, conditions, rights and priority are added.

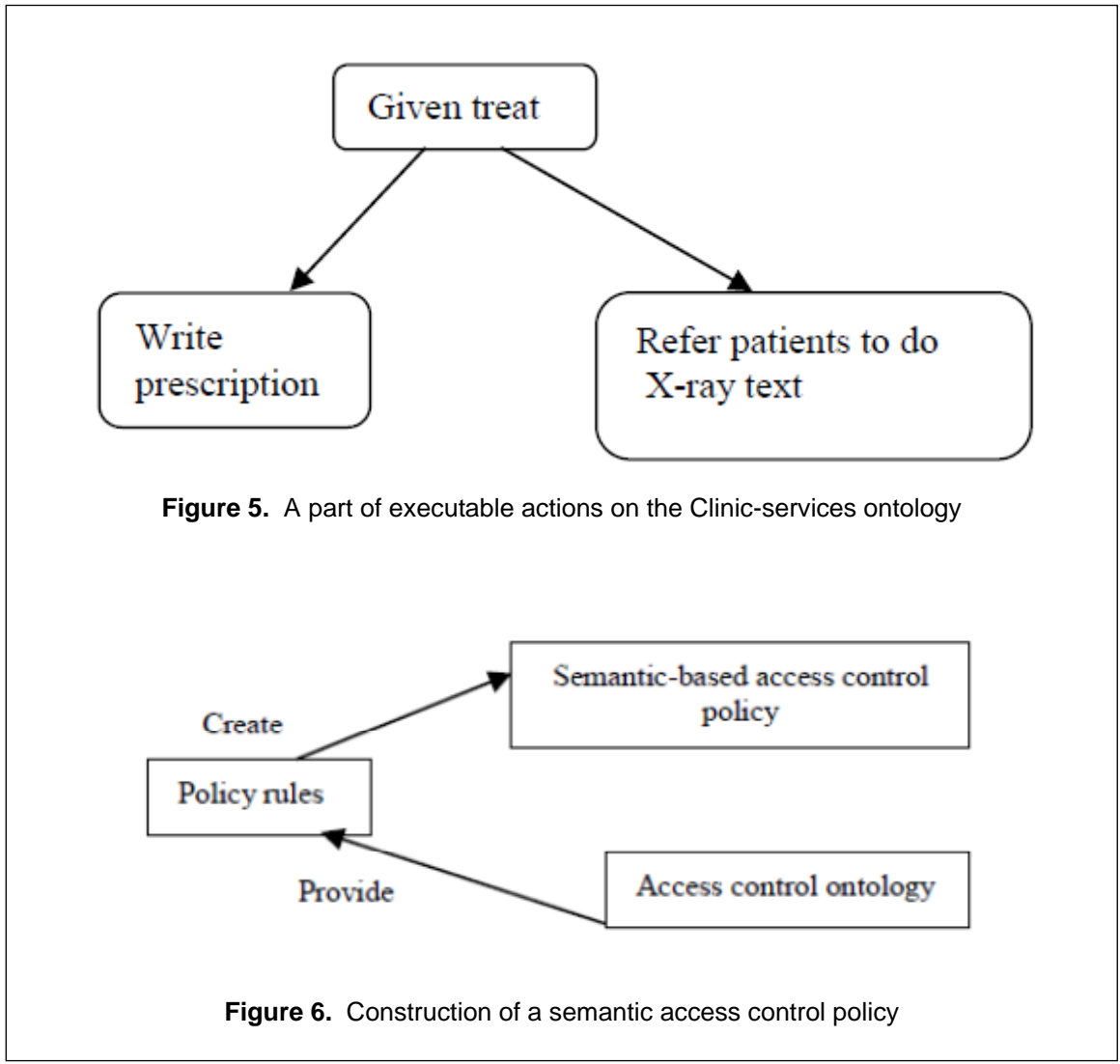

**Figure 5.** A part of executable actions on the Clinic-services ontology

**Figure 6.** Construction of a semantic access control policy

Policies are usually written in the form of rules. In the semantic access control, each policy must be associated with domain knowledge. Therefore, it is necessary to apply the domain knowledge to obtain the semantic elements for the semantic access control policy. The following figure depicts the relationships among an access control policy, policy rules, and access control ontology.

Semantic Policy Language (SPL) is based on the semantic properties about the resources to be accessed and about the attributes. The language can be applied to access control of cloud computing environments and the semantic access control is realized. An SPL policy is composed of a set of access rule elements. In SAC, the user poses a set of attributes, and the access control to resources is based on the specification of a set of attributes that the user has to present to access them [25]. Every access rule defines a particular combination of attributes required to gain access, associated with an optional set of actions (such as online permission) to be performed before access is granted. Figure 7 shows an example of a SPL policy requiring attributes for an authorized doctor. This policy includes one access rule indicating that access should be granted to all doctors authorized by the hospital administration authority. Any attributes that is proved equivalent to this one will be accepted because the equivalence attribute of the spl:

```
<?/xml version="1.0" encoding= "UTF-8"?>
</spl:policy>
 <spl:access_Rules>
   <spl:access_Rule Name = "Auth_doctors"  Public="false">
   <spl:attribute_Set>
    <spl:attribute attributeID = "Auth_doctors"  e = "Enabled">
    <spl:attribute_Name> doctor </spl:attribute_Name>
    <spl:SOA_ID> hospital_ADMIN </spl:SOA_ID>
   </spl:attribute_Set>
   </spl:access_Rule>
  </ spl:access_Rules>
</spl:policy>
```

**Figure 7.** Consulting_Access.xml Policy

attribute tag is set to "e=Enable". Furthermore, no information regarding the reason why the request is denied will be given to users that do not meet the access criteria because this access rule is not public (the public attribute of the spl: access_Rule tag is set to "false"). This feature is used by access control administrator to avoid unauthorized users learning about the existing access policies.

## 4. Semantic based access control in cloud computing

When a subject requests to perform an action on an object, the corresponding rules are evaluated by the enforcement engine for the request. A typical access control rule is expressed as a 3-tuple subject, object, action, such that a subject can perform some action on an object. In this article the semantic access control rule extends the typical access control rule 3-tuple to include purpose, condition, and right.

Each semantic access control rule defines a specific combination of: a purpose set, a condition set, rights of subjects and objects. The rule includes action, subject and object. Action specified operations including query, update, execution, or even more detailed action, such as read, write. The target are controlled by the rule is effective for all actions of the subject and object. The subject and the object can be a single user, a role or user group of ontology domain for support of SAC. Aside from the action to be performed, SAC permission explicitly states the intended purpose along with the conditions under which the permission can be given.

The rights define privileges that a subject can hold on an object. Rights require associations with subjects and objects. For example, if rights categories might be a "modification",  it means to change an existing digital object and create of a new object that reuses an original digital object.

A purpose describes the reasons for data collection and data access.  The purpose directly dictates how accesses to data items should be controlled.  Purposes are also widely used for specifying privacy rules in legislative acts and actual public policies [3, 22, 25]. Adopting purpose are the fundamental policies for private information concern with which data object is used for what purposes. For example, patients' age and email address are used for the purpose of medical analysis. And Alice checks some patient's blood pressure for the purpose of surgery.

The conditions define a logical expression for factors that the target may affect the applicability of the rules (such as state a condition requirement of resource). And the condition expression can be and/or relation several conditions. The conditions of the rules can be used in the semantic variables, which represent the attributes of subject, object, and action. It can effectively avoid the incomprehensibility problem caused by the heterogeneous of variables in the policy. If the conditions are empty, the rule applies in all circumstances without conditions. We illustrate through an example a permission expressed by purposes and conditions.

(1) "Clinic doctors can only access patients' email address if patients have given their consent."

```
<rule>
<Target>
<Subject name = "Anyperson" ontologyRef= "SO"/>
<AttributeVariable name = "doctors" type = "subject" ontologyRef=
"AtO">
<Object name = "Anyperson" ontologyRef= "OO"/>
<AttributeVariable name = "patients" type = "object" ontologyRef=
"AtO">
<Action name = "read" ontologyRef= "AO"/>
</Target>
<Right type = "modification"/>
<Purpose type = "treat"
<Condition type = "Equals" reference = "work more than three years"/>
</rule>
```

**Figure 8.** XML–based rule structure of SAC

(2) "Our clinic partners may access patients' information for research; patients' age must be over 50 years old."

In the first example, the condition = "patients have given their consent" and purpose = "n/a".

In the second example, the condition = "patients' age must be over 50 years old" and purpose = "research".

By modeling the access control domains using ontologies, SAC aims at considering semantic relationships with ontology to perform to make decision about an access request. Authorization base is a set of authorization rules in the form of $(s, o, a, pu, C, r)_{onto}$ in which $s$ is an entity in SO, $o$ is an entity defined in OO, and $a$ is an action defined in AO. In the other words, a rule determines whether a subject which presents a credential $s$ can have the access right $a$ on object $o$ or not.

We propose XML–based rule structure to support semantic policy description model based on semantic access control rules. We use an example to illustrate the basic element of semantic access control rules. In the example the policy is "Any doctor has greater than or equal to work more than three years can read the hospital's patient record, but it cannot be modified". The following example is described by XML-based rule structure in Figure 8.

# 5. Implementation with XACML

The implementation of the access control system focuses on using information stored in the semantic knowledge base. Figure 9 shows the architecture and message sequences of the implemented access control system referring back to the example in Figure 1.

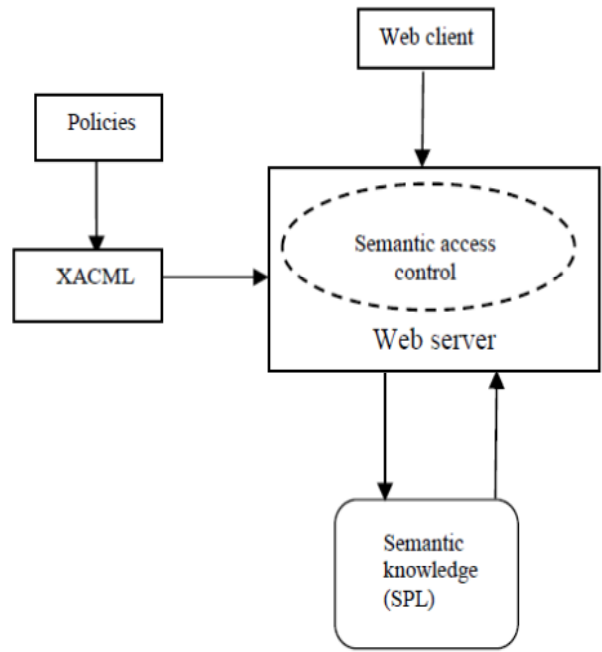

**Figure 9.** Access control system architecture

The message sequence works as follows:

1. A client presents an HTTP request to the web server.

2. At web server there is a semantic access control model with incoming requests to the server. The access control model performs queries to the semantic knowledge base in order to find attributes associated with subjects and objects.

3. Web server receives the attributes and translates the request to the XACML format.

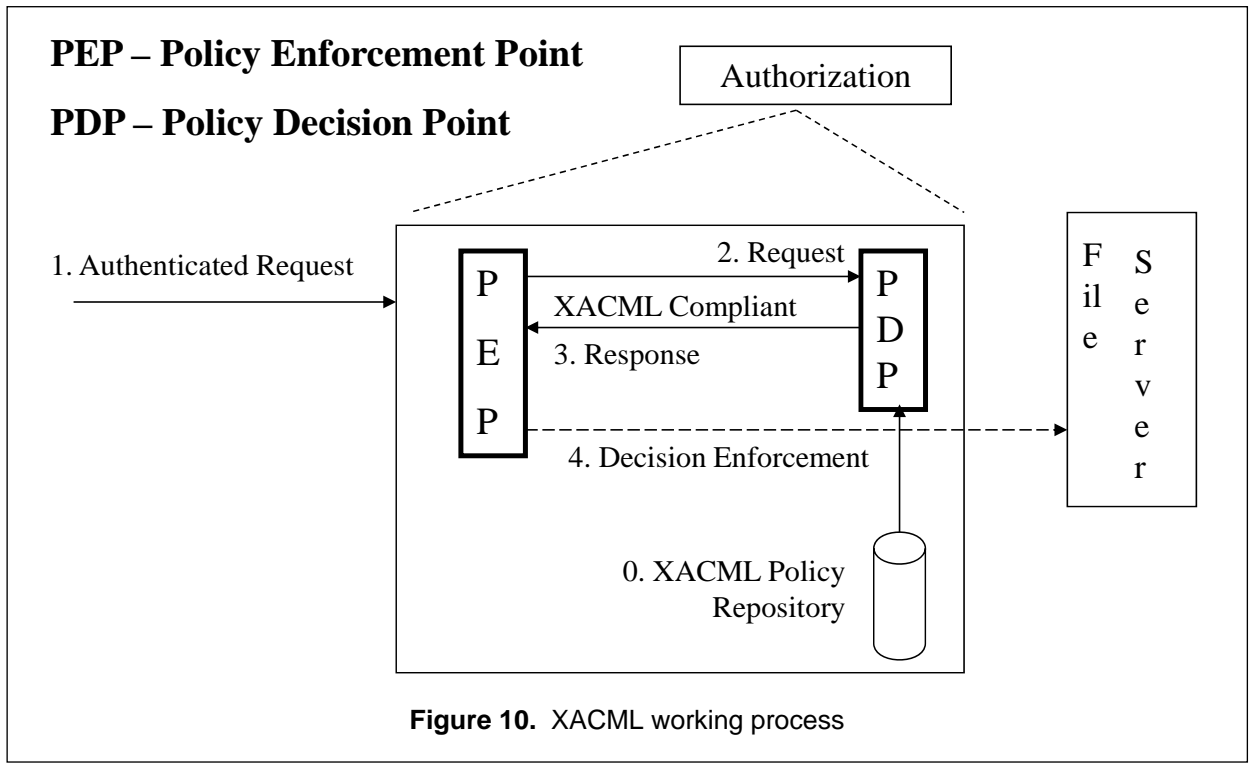

**PEP – Policy Enforcement Point**

**PDP – Policy Decision Point**

Authorization

1. Authenticated Request

P E P

2. Request

XACML Compliant

3. Response

P D P

4. Decision Enforcement

0. XACML Policy Repository

File Server

**Figure 10.** XACML working process

4. The XACML evaluates the request against an access control policy (SPL).

5. A response decision is sent back to the web server.

6. The web server has a check if the response is "permit"; the actual HTTP request is forwarded to the web server. Otherwise, an appropriate HTTP response is sent to the web client with an error message.

XACML describes both a policy language and an access control decision request/response language (both written in XML). The policy language is used to describe general access control requirements, and has standard extension points for defining new functions, data types, combining logic, etc. The request/response language forms a query to check whether or not a request should be allowed, and interprets the result. The response always includes an answer about whether the request should be allowed using one of four values: Permit, Deny, Indeterminate or Not Applicable.

Assuming that someone wants to take some action on a resource. They will make a request to whatever actually protects that resource (like a filesystem or a web server), which is called a Policy Enforcement Point (PEP). The PEP will form a request based on the requester's attributes, the resource in question, the action, and other information pertaining to the request. The PEP will then send this request to a Policy Decision Point (PDP), which will look at the request and some policy that applies to the request, and come up with an answer about whether access should be granted. That answer is returned to the PEP, which can then allow or deny access to the requester. Note that the PEP and PDP might both be contained within a single application, or might be distributed across several servers. In addition to providing request/response and policy languages, XACML also provides the other pieces of this relationship, namely finding a policy that applies to a given request and evaluating the request against that policy to come up with a yes or no answer. XACML works as described in Figure 10.

## 6. Related work

Our work is related to many areas of privacy preserving access control, especially private data management in e-Healthcare system. We also exploit the tremendous work carried out for semantic access control which mainly focuses on secure management of data in e-Healthcare.

Role-based access control is commonly accepted as the most appropriate paradigm for the implementation of access control in complex scenarios. Reid et al. [17] presented that RBAC has received considerable attention in the context of health care, particularly in the hospital environment. With these models, the roles are organized hierarchically and the specialized roles inherit the privileges of the more general roles. If certain privilege is assigned to an employee role, possession of any of the superior roles enables the same privilege. However, the structure of groups in RBAC is usually assumed static; it is not flexible enough to cope with the requirements of more dynamic systems. On the other hand, RBAC is that the mechanisms are built on three components: "user", "role" and "group". Roles and groups can facilitate management in corporate information systems.

In most cases we need new resources which are incorporated to the system continuously and each resource may possibly need a different group structure and access control policy. Other traditional access control schemes such as Mandatory Access Control (MAC), Discretionary Access Control (DAC) are not appropriate for the system with a very large number of registered users. By considering attributes and access control ontology system to the basis of the access control model, the Semantic access control (SAC) model [28, 32] provides an appropriate solution for large environments such as cloud computing. The SAC model has been implemented on the basis of the Semantic Policy Language (SPL) to specify the access control criteria, and the semantic integration of an external authorization entity. Our approach has more powerful and suitable to implement mutual understanding and semantic interoperability of distributed policy in cloud computing environments.

Extensible Access Control Markup Language (XACML) [15] is a standard access control policy description language. XACML can be applied to represent the functionalities of most policy representation mechanisms and express access control statements [17]. Damiani et al. [6] extended the XACML by adding the capability to designate subjects and objects via generic RDF statements. Priebe et al. [20] extend the XACML architecture with an ontology-based inference facility for attributes management and mapping. Because these approaches further complicate the access control process of XACML and XACML has some features provided by those languages are not appropriate in WS scenarios.

# 7. Conclusions and future work

In this article we have proposed the semantic approach of SAC is the foundation to achieve access control in cloud computing environment. The SAC model is scalable, applicable to different environment and covers other access control models. SAC extends RBAC by considering the semantics of objects and associates permission with concepts instead of objects. Ontology is used for cloud computing environment with highly heterogeneous and structured vocabularies. Considering the limitations of traditional access control method in the cloud computing, this paper introduces the semantic web technologies to the distributed role-based access control method and an ontology-based semantic access control in e-Healthcare system. In the SAC, we use some syntax elements, such as subjects, objects based on attributes and action; and we add more elements purposes, conditions, rights and priority in the SAC model. This approach can easier solve the problem of access in heterogeneous, distributed and large environments and ideal for doing the cross organizational work in cloud computing environment. We are also working on the implement of this approach in e-Healthcare applications.